\documentclass[prb,aps,showpacs,unsortedaddress,superscriptaddress,amsmath,amssymb,floatfix,twocolumn,showkeys]{revtex4-1}
\usepackage{graphicx}
\usepackage{dcolumn}
\usepackage{bm}
\usepackage[hidelinks,colorlinks=true,linkcolor=blue,citecolor=blue]{hyperref}
\usepackage[mathlines]{lineno}
\usepackage{braket}
\usepackage[usenames]{color}
\usepackage{float}
\usepackage{amsmath}
\usepackage{subcaption}
\begin{document}
\title{Electronic properties of magnetic semiconductor $\textrm{CuMnO}_{2}$ : a first principles study}
\author{Apurba Sarkar}
\affiliation{Department of Physics, Kazi Nazrul University, Asansol-713340,West Bengal, India}
\author{Joydeep Chatterjee}
\affiliation{Department of Physics, Indian Institute of Technology Kharagpur, Kharagpur-721302, West Bengal India}
\author{Soumya Mukherjee}
\affiliation{Department of Metallurgical Engineering, School of Mines and Metallurgy, Kazi Nazrul University, Asansol-713340, West Bengal, India}
\author{Arghya Taraphder}
\affiliation{Department of Physics, Indian Institute of Technology Kharagpur, Kharagpur-721302, West Bengal India}
\author{Nandan Pakhira}
\affiliation{Department of Physics, Kazi Nazrul University, Asansol-713340,West Bengal, India}
\begin{abstract}
Geometrically frustrated magnetic semiconductor $\textrm{CuMnO}_{2}$ has potential applications as photo-catalyst, in photochemical cells and multi-ferroic devices. Electronic band structure in the antiferromagnetic and 
ferromagnetic phases of $\textrm{CuMnO}_{2}$ were calculated using first principle density functional theory (DFT) as implemented in VASP. Electronic band structure in the antiferromagnetic state shows indirect band gap 
($\sim 0.53$ eV) where as in the ferromagnetic state it shows half-metallic state with 100\% spin polarization. The half-metallic state arises due to \textit{double exchange} mechanism. In the half-metallic state the density 
of states for the up spin channel shows asymmetric power law behaviour near the Fermi level while the down spin channel shows fully gapped behaviour. The calculated magnetic moment of Mn atom in the ferromagnetic 
(3.70 $\mu_{B}$) and antiferromagnetic (3.57 $\mu_{B}$) states are consistent with experimental values. Our calculation predicts potential application of $\textrm{CuMnO}_{2}$ in spintronic devices especially in the 
ferromagnetic state, as a spin injector for spin valves in spintronic devices.
\end{abstract}
\maketitle
\section{Introduction}
\begin{figure}[h]
\centering
\includegraphics[scale=0.3]{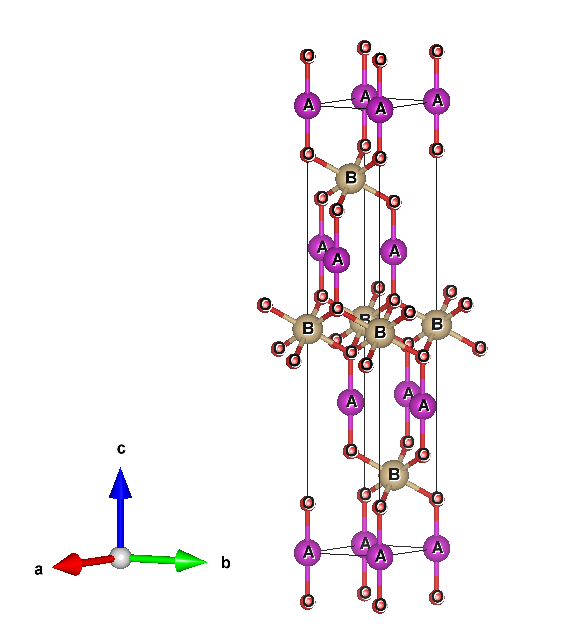}
\caption{Conventional unit cell of rhombohedral (space group R$\overline{3}$m) delafossite compound $\textrm{ABO}_{2}$ (A= Cu, Ag; B= Fe, Cr, Mn, Co, Ni). The structure consists of layers of slightly distorted octahedral 
$\textrm{BO}^{6}$  separated by the O–A–O linear linkage. }
\label{Fig:1}
\end{figure}

Magnetic semiconductors such as full Heusler alloys~\cite{ali2021crystal}, half Heusler alloys~\cite{benmakhlouf2018structural,tariq2023first}, diluted magnetic semiconductors (DMS)~\cite{ismail2021study} etc. have been 
commercially used for the fabrication of spintronic devices. Geometrically frustrated magnetic semiconductors with chemical formula $\textrm{ABO}_{2}$ (A = Cu, Ag ; B = Fe, Cr, Mn, Co, Ni) have attracted a lot of attention 
due to their diverse and fascinating electronic and magnetic properties as well as their potential technological applications such as multi-ferroic oxide materials, photovoltaic and photo-catalytic semiconductors, thermo-
electric materials etc.~\cite{amrute2013solid,schorne2019insights,sullivan2016copper,xiong2018polyvinylpyrrolidone}. Some of the materials (like, $\textrm{CuFeO}_{2}$, $\textrm{CuCrO}_{2}$, $\textrm{NaCoO}_{2}$)
~\cite{takahashi2003single,seki2008spin,haraldsen2010multiferroic}) showing triangular-lattice antiferromagnetic (TLA) state have layered delafossite structure as shown in Fig.~\ref{Fig:1}, where the monovalent $A^{+}$ 
cations are linearly coordinated with two $\textrm{O}^{-2}$ ions, forming O-A-O dumbbells and trivalent $\textrm{B}^{3+}$ cations are surrounded by six $\textrm{O}^{-2}$ ions, forming $\textrm{BO}^{6}$ edge-sharing-octahedral 
layers. These two layers are stacked along the $c$-axis. The two-layer repeating structure belongs to the space group P63/mmc with 2H hexagonal symmetry while the three layer repeating structure belongs to the space group 
R$\overline{3}$m with 3R rhombohedral symmetry. These class of semiconducting magnetic materials show giant magneto-resistance (GMR), tunnel magneto-resistance (TMR) as well as anisotropic magneto-resistance (AMR)
~\cite{wang2019giant,ismail2021study}.  

Another example of geometrically frustrated magnetic system is $\textrm{CuMnO}_{2}$ (crednerite). $\textrm{CuMnO}_{2}$ is chemically stable (over pH range 7-14), transparent $p$-type semiconductor with direct band gap 
in the visible region and hence absorbs a large part of sunlight. This material has potential application as photo-catalyst, in photochemical cells, multiferroic devices, etc.~\cite{zorko2015magnetic,poienar2018microwave,
kato2009oxygen}. In previous studies~\cite{damay2009spin,ushakov2014orbital} it has been shown that $\textrm{CuMnO}_{2}$ exhibits two types of magnetic ordering. At low temperatures, ($T < T_{N}$ = 65K) it shows three-
dimensional antiferromagnetic (AFM) spin ordering where as ferromagnetic (FM) spin ordering at high temperatures and behaves like metal~\cite{ushakov2014orbital}. On the other hand, AFM $\textrm{CuMnO}_{2}$ is insulating 
with band gap of 0.2 eV~\cite{hiraga2009optical}. It is interesting to mention that ferromagnetic semiconductors with strong spin polarization can be used as spin injector for spin valves in spintronic devices. 
Antiferromagnetic semiconductors like bulk $\textrm{SrIr}_{2}\textrm{O}_{4}$~\cite{FinaNatComm2014}, Cr-Doped Ge-Core/Si-Shell nanowires~\cite{AryalNanoLett2021}, porous boron nitride ($\textrm{B}_{6}\textrm{N}_{6}$) 
sheet~\cite{AbdullahiCompTheoChem2021}, 2D monolayers of MnSi2N4~\cite{ChenFrontChem2022} have many potential applications in future spintronic devices. In this context, geometrically frustrated  $\textrm{CuMnO}_{2}$, 
because of its unique magnetic properties, is one of the most promising material for spintronic devices.   

To the best of our knowledge there is no reported theoretical study on the electronic structure of spin-polarized $\textrm{CuMnO}_{2}$. In this work, using first principle density functional theory 
(DFT)~\cite{orio2009density}, we have investigated the electronic structures and density of states for both the antiferromagnetic (AFM) and ferromagnetic (FM) states of $\textrm{CuMnO}_{2}$. We find that the 
anti-ferromagnetic state is a small bandgap semiconductor while the ferromagnetic state is a half-metal with 100\% spin polarization. It is interesting to mention that, Ismail et. al.~\cite{ismail2023ab}, using first 
principles DFT, have reported half-metallic behavior of a new class of delafossite material with chemical formula $\textrm{SMoO}_{2}$ (S= Na, K, Rb, Cs). The organization of the rest of the paper is as follows : in 
Sec. II we discuss about the crystal structure of $\textrm{CuMnO}_{2}$, in Sec. III we provide the computational details. Then in Sec. IV we discuss our results for both type of magnetically ordered states, in Sec. V we 
discuss about the magnetic properties of the system in both the phases and finally in Sec. VI we conclude.

\section{Crystal Structure}
$\textrm{CuMnO}_{2}$ has monoclinic structure belonging to the space group C2/m. The monoclinic C2/m structure instead of the usual rhombohedral R$\overline{3}$m structure arises due to lattice distortion in the presence 
of Jahn-Teller effect on the $\textrm{Mn}^{3+}$ ions in the crystal-field split $\textrm{e}_{g}$ orbital manifold of $\textrm{d}^{4}$ electronic configuration ($\textrm{t}_{2g}^{3}$$\textrm{e}_{g}^{1}$). The optimized 
crystal structure of $\textrm{CuMnO}_{2}$ is shown in Figure 1 (a) and (b). In this structure alternating $\textrm{MnO}^{6}$ octahedral layers are separated by O-Cu-O layer along the $c$-axis. There are two short and 
four long distances for each Mn atom which significantly effects the magnetic arrangement of $\textrm{Mn}^{3+}$ cations and frustrates the system. In each unit cell there are 2 Cu, 2 Mn and 4 O atoms. In Table~\ref{tab1} 
we show the the Wyckoff positions of various atoms. The system is topologically representative of a frustrated square lattice with nearest neighbour ($J_{1}$) and next nearest neighbour ($J_{2}$) interactions. This 
frustration gets relieved through a magneto-elastic coupling to the lattice as evidenced by a structural phase transition, associated with magnetic ordering, from a monoclinic ($C2/m$) to a strained triclinic $C\bar{1}$ 
phase.   
\begin{table}[htbp]
\centering
	\caption{\small{Atomic positions of various elements}}\label{tab1}%
\begin{tabular}{|c|c|c|c|c|}
\hline
	Wykcoff & element & x & y & z\\
\hline
           2a   &    Mn    & 0 & 0 & 0 \\
\hline
           2d   &    Cu   & 0 & 0.5 & 0.5 \\
\hline
           4i   &    O    & 0.904961 & 0.5 & 0.179972 \\
\hline
\end{tabular}
\end{table}	
\begin{figure}[h]
	\centering
	\begin{subfigure}{0.2\textwidth}
		\centering
		\includegraphics[scale=0.17]{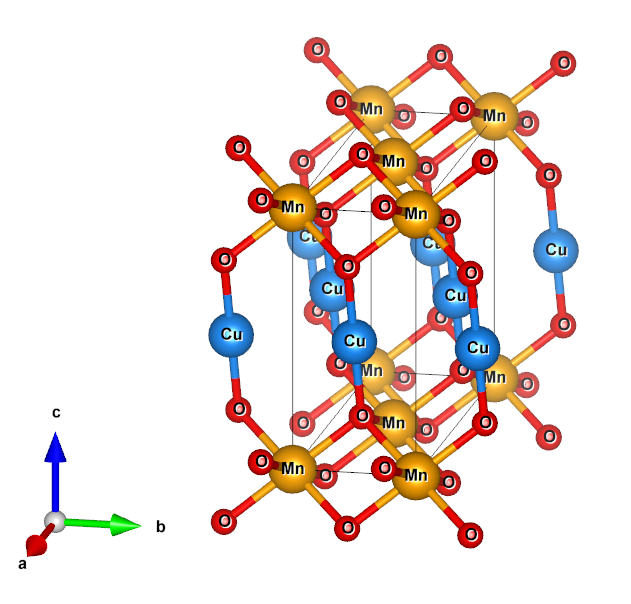}
		\caption{}
	\end{subfigure}%
	\begin{subfigure}{0.2\textwidth}
		\centering
		\includegraphics[scale=0.17]{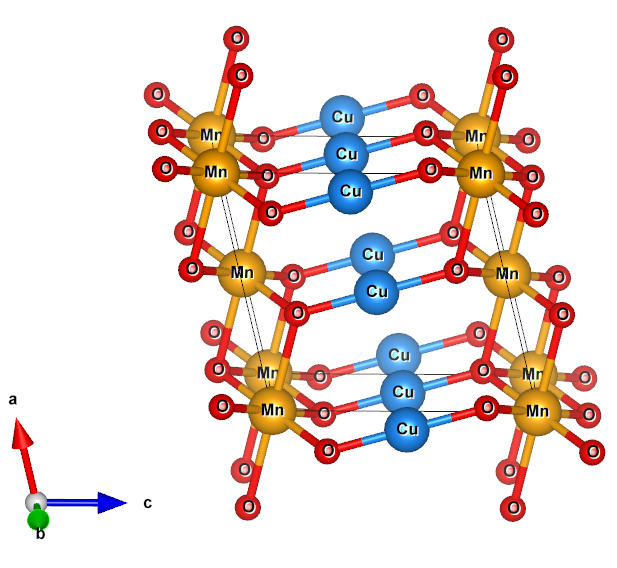}
		\caption{}
	\end{subfigure}
	\caption{(a) and (b) are the standard and side view of the optimized crystal structure of $\textrm{CuMnO}_{2}$ (Space group C2/m).}
	\label{crystal structure}
\end{figure}

\section{Computational Details}
All the calculations were performed using the density functional theory (DFT) framework, as implemented in the VASP code~\cite{hafner2008ab}. To treat the exchange and correlations effect between core and valence 
electrons, we used the Perdew-Burke-Ernzerhof (PBE)~\cite{ernzerhof1999assessment} exchange-correlation (XC) functionals in the frame work of generalized gradient approximation (GGA)~\cite{perdew1996generalized} and 
projector augmented wave (PAW)~\cite{blochl1994projector} basis. The experimental lattice parameters and atomic positions of CuMnO2 were taken from material project website. The constituent atoms and their valance state 
are: $\textrm{Cu : [Ar]} 3d^{10}4s^{1} $, $\textrm{Mn : [Ar]} 3d^{4}$ and $\textrm{O :} 1s^{2} 2s^{2} 2p^{6}$ , respectively. The atomic positions, lattice vectors, cell angles and the cell volume were allowed to relax. 
The optimized lattice parameters along with their experimental values are summarized in Table~\ref{tab2}. The convergence criterion for energy and force were set to $10^{-8}$ eV and 0.001 eV/\AA, respectively. We have 
used an energy-cutoff 520 eV and a $15\times15\times15$ Monkhorst-Pack~\cite{monkhorst1976special} $k$-mesh for structural optimization and GGA selfconsistent calculation while for the density of states (DOS) calculation,
we have used a denser Monkhorst-Pack $25\times25\times25$ $k$-mesh. The Brillouin-zone (BZ) integration was performed on $\Gamma$ centered symmetry reduced Monkhorst-Pack meshes using a Gaussian smearing with $\sigma$ = 
0.05 eV. The electronic band structure was plotted along the paths involving high symmetry points $\Gamma$, $C$, $C_{2}$, $Y_{2}$, $M_{2}$, $D$, $D_{2}$, $A$, $V_{2}$. Calculated bands were plotted along the high 
symmetry directions $\Gamma - C$, $C_{2}-Y_{2}-\Gamma-M_{2}-D$, $D_{2}-A-\Gamma - V_{2}$. Crystal structure visualization and analysis were performed using the VESTA~\cite{momma2011vesta} software. GGA+U calculations were 
also carried out with on-site Coulomb repulsion term. We used $U_{\textrm{eff}}= U-J_{H}=4.1$ eV for Mn atoms which is consistent with the previously reported values~\cite{ushakov2014orbital}.

\begin{table}[h]
\caption{Comparison between experimental and calculated (this work) lattice constants of $\textrm{CuMnO}_{2}$ under GGA approximation.}\label{tab2}%
         \begin{tabular}{@{}|l|l|l|l|l|l|l|}
\hline
		 \multicolumn{1}{|c|}{} &\multicolumn{3}{|c|}{Experimental}&\multicolumn{3}{|c|}{GGA} \\
\hline
		System& $a$(\AA) & $b$(\AA) & $c$(\AA) & $a$(\AA) & $b$(\AA) & $c$(\AA) \\
		\hline
			$\textrm{CuMnO}_{2}$ & 5.58 & 2.88 & 5.89 & 5.68 & 2.89 & 5.90 \\
		\hline	
\end{tabular}
\end{table}

\section{Results and Discussions}
In Table~\ref{tab2} we compare the lattice constants obtained from structural relaxation under GGA against their observed experimental values. As can be clearly observed, the calculated values are very close to the 
observed experimental values. This validates the computed relaxed structures against their observed lattice structure. From the relaxed structure we calculated the self-consistent field and subsequently the electronic 
band structure and density of states. We consider both type of magnetically ordered states.
\subsection{Electronic structure of Antiferromagnetic $\textrm{CuMnO}_{2}$} 
We first consider the antiferromagnetically ordered state. According to Damay et. al.~\cite{damay2009spin} the low temperature magnetic order with propagation vector $\mathbf{k} = (-\frac{1}{2},\frac{1}{2},\frac{1}{2})$ 
is described as spins antiferromagnetically aligned along [1 -1 0] and ferromagnetically aligned along [1 1 0], corresponding to collinear antiferromagnetic order predicted for a frustrated two-dimensional 
antiferromagnetic Heisenberg model on a square lattice for $J_{2}/J_{1} > 0.5$ in the presence of spin-lattice coupling. 

\begin{figure}[htbp]
	\centering
	\includegraphics[scale=0.2]{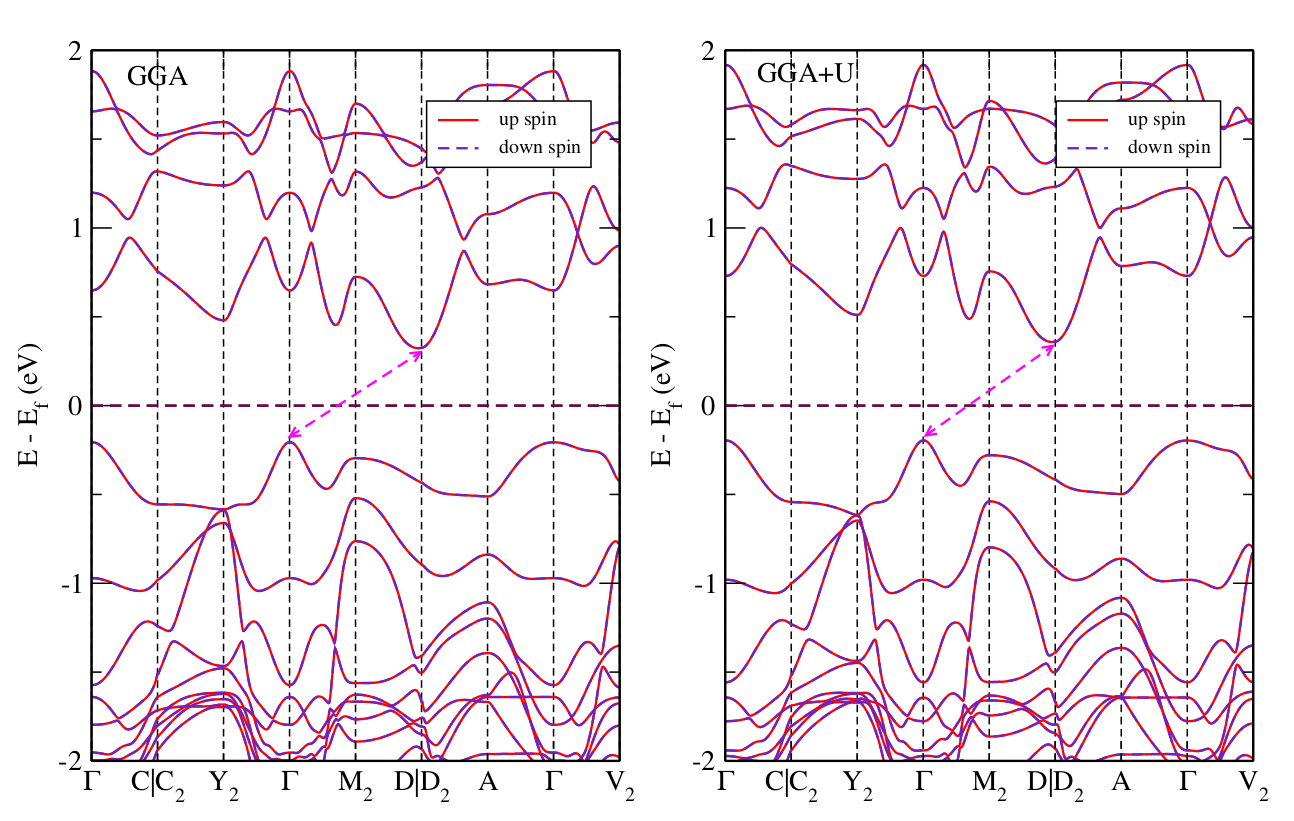}
	\caption{Electronic band structure in the antiferromagnetic state of $\textrm{CuMnO}_{2}$, calculated under GGA [panel (a)] and GGA+U [panel (b)] approximations. The Fermi energy is set to 0 eV.}
	\label{Fig:AFMbandStructure}
\end{figure}

In Fig.~\ref{Fig:AFMbandStructure} we show the electronic band structure of $\textrm{CuMnO}_{2}$ in the antiferromagnetic state, calculated under GGA and GGA+U approximations. As stated earlier, the bands are plotted 
along various high symmetry directions and the Fermi level is set to zero. The calculated band structure shows that $\textrm{CuMnO}_{2}$ in the antiferromagnetic state is an indirect semiconductor with band gap 0.53 eV. 
The valence band maximum (VBM) is situated at the $\Gamma$ point while the conduction band minimum (CBM) is located at the $D$ point and is indicated by the dotted arrow. Under GGA+U approximation the indirect band gap 
increases by 0.02 eV. In order to understand the characteristics and detailed orbital structure, we also have calculated the total as well as partial density of states for various atoms. 
\begin{figure}[htbp]
	\centering
	\includegraphics[scale=0.35]{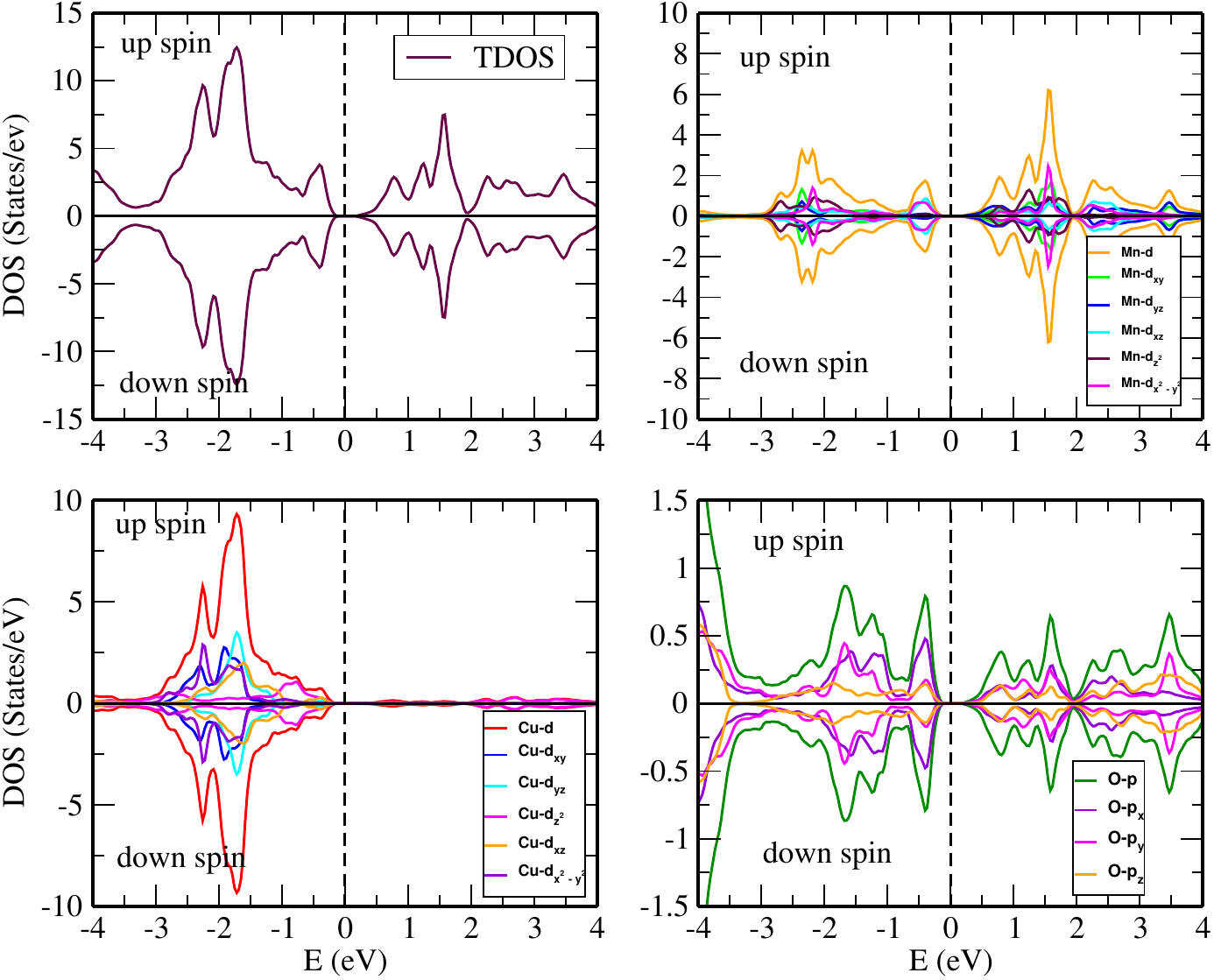}
	\caption{Total and partial density of states of $\textrm{CuMnO}_{2}$ in the anti-ferromagnetic state under GGA approximations. The Fermi level is set to 0 eV. }
	\label{Fig:AFMdosGGA}
\end{figure}
\begin{figure}[htbp]
	\centering
	\includegraphics[scale=0.35]{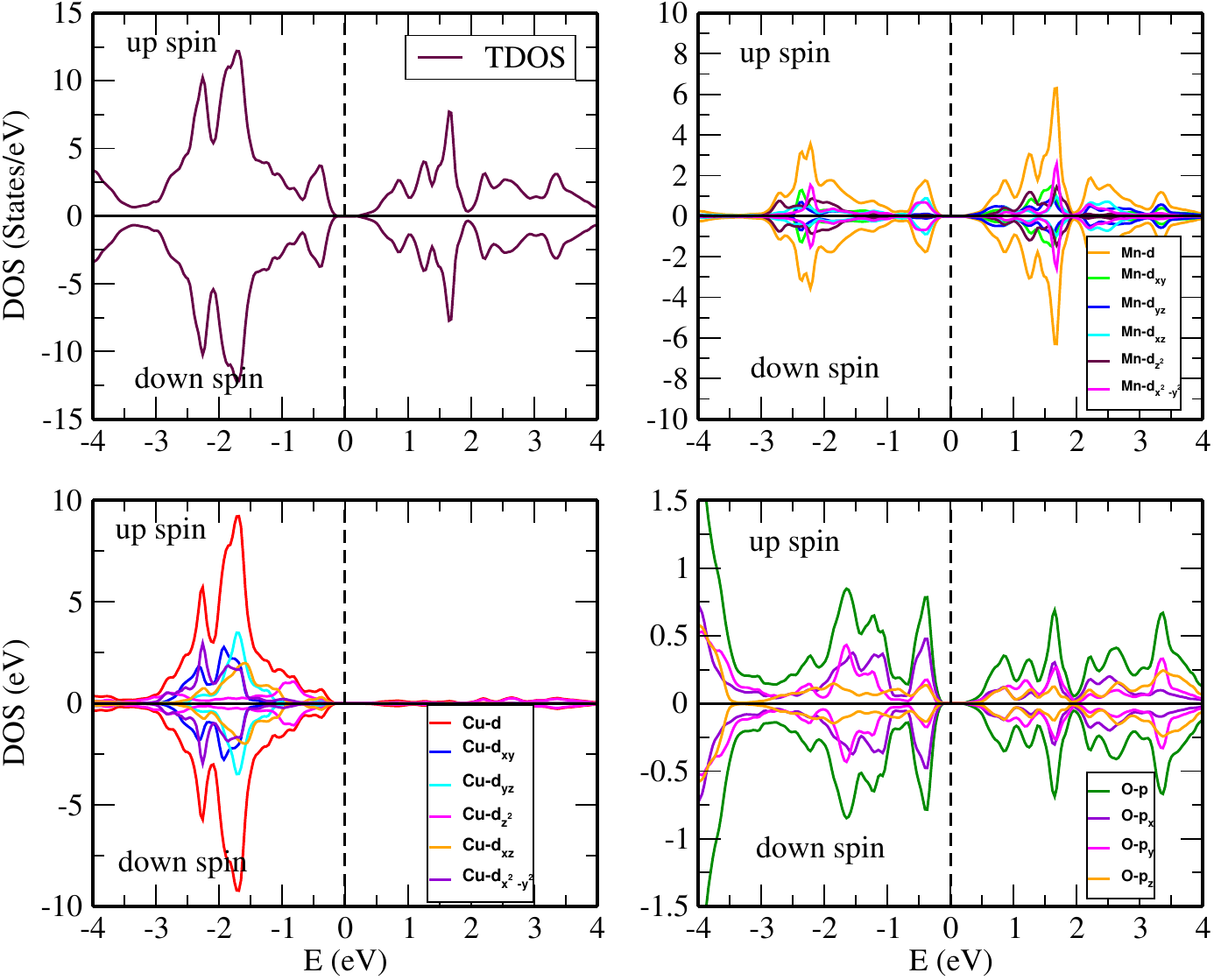}
	\caption{Total and partial density of states of $\textrm{CuMnO}_{2}$ in the anti-ferromagnetic state under GGA+U approximations. The Fermi lelvel is set to 0 eV. }
	\label{Fig:AFMdosGGA_U}
\end{figure}
In Fig.~\ref{Fig:AFMdosGGA} and Fig.~\ref{Fig:AFMdosGGA_U} we have shown the total density of states (TDOS) as well as the partial density of states (PDOS) of each atom in the low temperature antiferromagnetic phase 
under GGA and GGA+U approximations, respectively . Further, we also have shown orbital resolved partial density of states for each atom in a given unit cell. The total as well as partial density of states show a clear 
gap at the Fermi level, consistent with the semiconducting nature of the antiferromagnetic state. The density of states away from the Fermi level is dominated by broad peaks arising due to Mn-$3d$ and Cu-$3d$ orbitals. 
In particular the braod peak at -0.4 eV corresponding to the VB arises mainly due to hybridization between Mn $3d_{x^2-y^2}$, $3d_{xz}$ and O $2p_{x}$, $2p_{y}$ and $2p_{z}$ orbitals. This is expected as in $MnO_{6}$ 
octahedra each $Mn^{3+}$ cation is in $d^{4}(t^{3}_{2g}e^{1}_{g})$ state. The VB as well as antiferromagnetic ordering arises due to overlap between Mn $3d_{x^2-y^2}$ and O $2p_{x}$, $2p_{y}$  orbital along each Mn-O-Mn 
bond giving rise to antiferromagnetic $e_{g}-e_{g}$ superexchange between half-filled $e_{g}$ orbitals. Also two Mn $3d_{xz}$ orbitals of two neighbouring $MnO_{6}$ octahedra can overlap via O $2p_{z}$ orbitals giving 
rise to antiferromagnetic $t_{2g}-t_{2g}$ superexchange. The broad spectral fetures in the range -4 eV to -1 eV arises mainly due to Cu $3d$ orbitals present in each O-Cu-O layer along the $c$-axis. The conduction band 
arises due to transition between various hybridized Mn-$3d$ and O-$2p$ orbitals. The conduction band peaks between +1 eV to +2 eV are much sharper than the peaks of the valance band. The origin of these peaks are due to 
transitions between crystal field split Mn-$3d$ orbitals which maintains very strong atomic character. The conduction band minimum is mainly composed of Mn-$3d$ states with the minor contribution of O-$p$ states 
indicating some Mn-O interactions. The symmetric nature of the TDOS for both the up and down spin channel confirms the antiferromagnetic nature of the ground state.

\subsection{Electronic structure of Ferromagnetic $\textrm{CuMnO}_{2}$} 
\begin{figure}[htbp]
	\centering
	\begin{subfigure}[b]{0.45\textwidth}
	\includegraphics[width=1.0\textwidth]{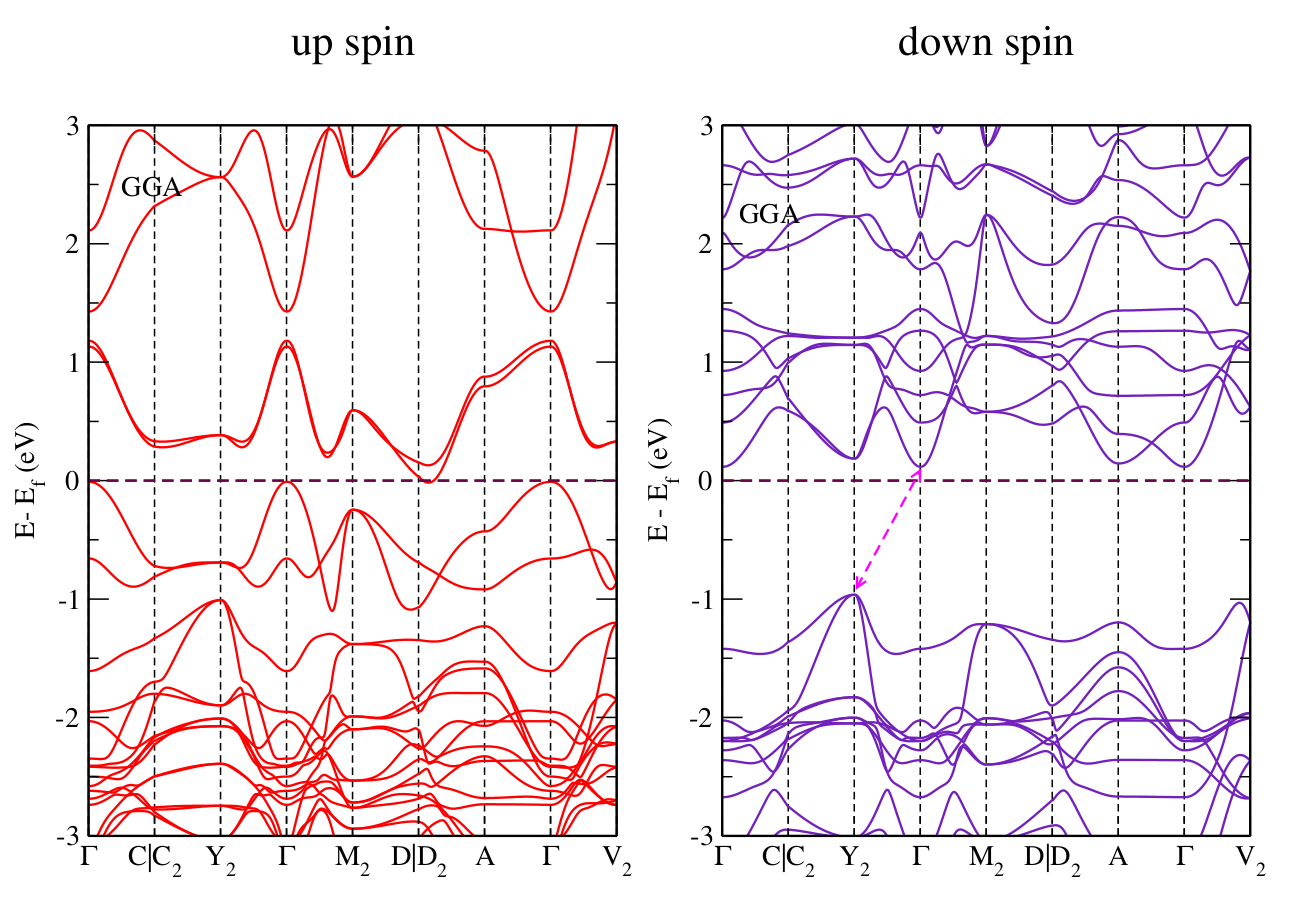}
	\caption{}
	\label{bands}
    \end{subfigure}
	\begin{subfigure}[b]{0.45\textwidth}
	\centering
	\includegraphics[width=1.0\textwidth]{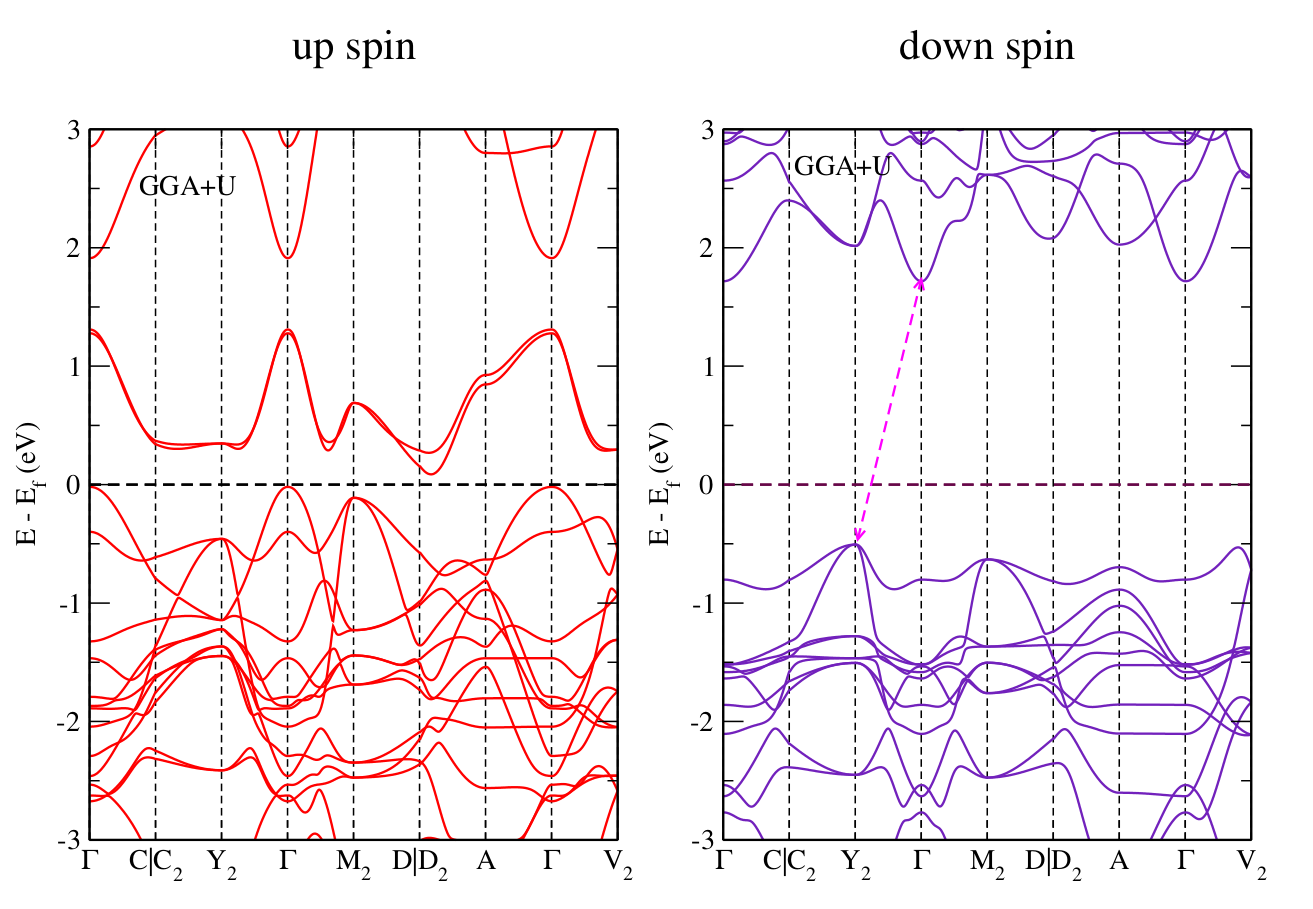}
	\caption{}
	\label{band}
	\end{subfigure}
	\caption{Electronic band structure in the ferromagnetic state of $\textrm{CuMnO}_{2}$, calculated under GGA [panel (a)] and GGA+U [panel (b)] approximations. The top of the valence band is set to 0 eV.}
	\label{Fig:bandsFM}
\end{figure}

Next we consider the ferromagnetic state of $\textrm{CuMnO}_{2}$. It has been reported~\cite{damay2009spin,ushakov2014orbital,VecchiniPRB2010,PoienarCM2011} that the high temperature state of $\textrm{CuMnO}_{2}$ is ferromagnetic in nature. 
In Fig.~\ref{Fig:bandsFM} we have shown spin-polarized electronic band structure in the ferromagnetic state of $\textrm{CuMnO}_{2}$ under GGA [panel (a)] and GGA+U [panel (b)] approximations. As can be clearly 
observed electronic bands corresponding to the down spin channel shows semiconducting nature while for the up spin channel some bands touch the Fermi level and hence shows the metallic characteristic for up-spin channel. Therefore $\textrm{CuMnO}_{2}$ would behave like  half-metallic ferromagnetic material with 100\% spin polarization. It is important to mention that half-metallic ferro-magnets (HMFM) is essential for spintronic 
devices~\cite{WolfScience,PickettPhysicsToday}. In particular $\textrm{CuMnO}_{2}$ can be used as a spin injector for spin valves in opto-spintronic devices. In the down-spin channel, valence band maxima (VBM) lies at 
$\textrm{Y}_{2}$ point and  conduction band minima (CBM) lie at $\Gamma$ point which indicates the indirect band gap semiconducting behavior of $\textrm{CuMnO}_{2}$ for down-spin channel. The indirect band gaps in the 
down-spin channel (along $\textrm{Y}_{2}$ - $\Gamma$ symmetry), obtained under GGA and GGA+U approximations are found to be 1.08 eV and 2.22 eV, respectively. 

The half-metallic behavior of $\textrm{CuMnO}_{2}$ arises due to the hybridization process that occurs between O-$2p$ states and Mn-$3d$ states. The hybridization mainly occurs through the \textit{double exchange} 
mechanism~\cite{saini2019prediction}. As already mentioned, in each $\textrm{MnO}_{6}$ octahedra $Mn^{+3}$ ions are in the $t_{2g}^{3}e_{g}^{1}$ configurations. The 3 electrons in the $t_{2g}$ manifold effectively forms 
a core spin with $S=3/2$. According to Anderson-Hasegawa model~\cite{AndersonPR1955} for double exchange electrons in singly occupied $e_{g}$ manifold is strongly coupled with $t_{2g}$ core spin through Hund's coupling 
$J_{H}$. In the limit $J_{H}\rightarrow\infty$ the spin of the $e_{g}$ electron gets effectively locked with the core spin and the effective hopping amplitude between two Mn atoms (via O atoms) gets reduced by a factor 
$\cos(\theta/2)$, where $\theta$ is the angle between two core spins. Under this scenario conduction through only one spin channel is allowed and the bands corresponding to other spin channel gets completely integrated 
out. For finite $J_{H}$ the energy bands corresponding to the unfavourable spin channel gets gapped out at the Fermi level and the energy gap is proportional to $J_{H}$. 

The observed features of the electronic band structure can be easily explained in terms of the scenario presented above. In the ferromagnetic phase the spins were in the up configuration there by making the up spin 
channel favourable and the down spin channel unfavourable. This will give rise to 100\% spin polarized transport through up spin channel while the down channel will be gapped out. The energy gap will be directly 
proportional to effective Hund's coupling strength. In the presence of $U$ [under GGA+U approximation] the gap further enhances due to Coulomb energy cost for double occupancy.  
\begin{figure}[htbp]
	\centering
	\begin{subfigure}[b]{0.5\textwidth}
	\includegraphics[width=1.0\textwidth]{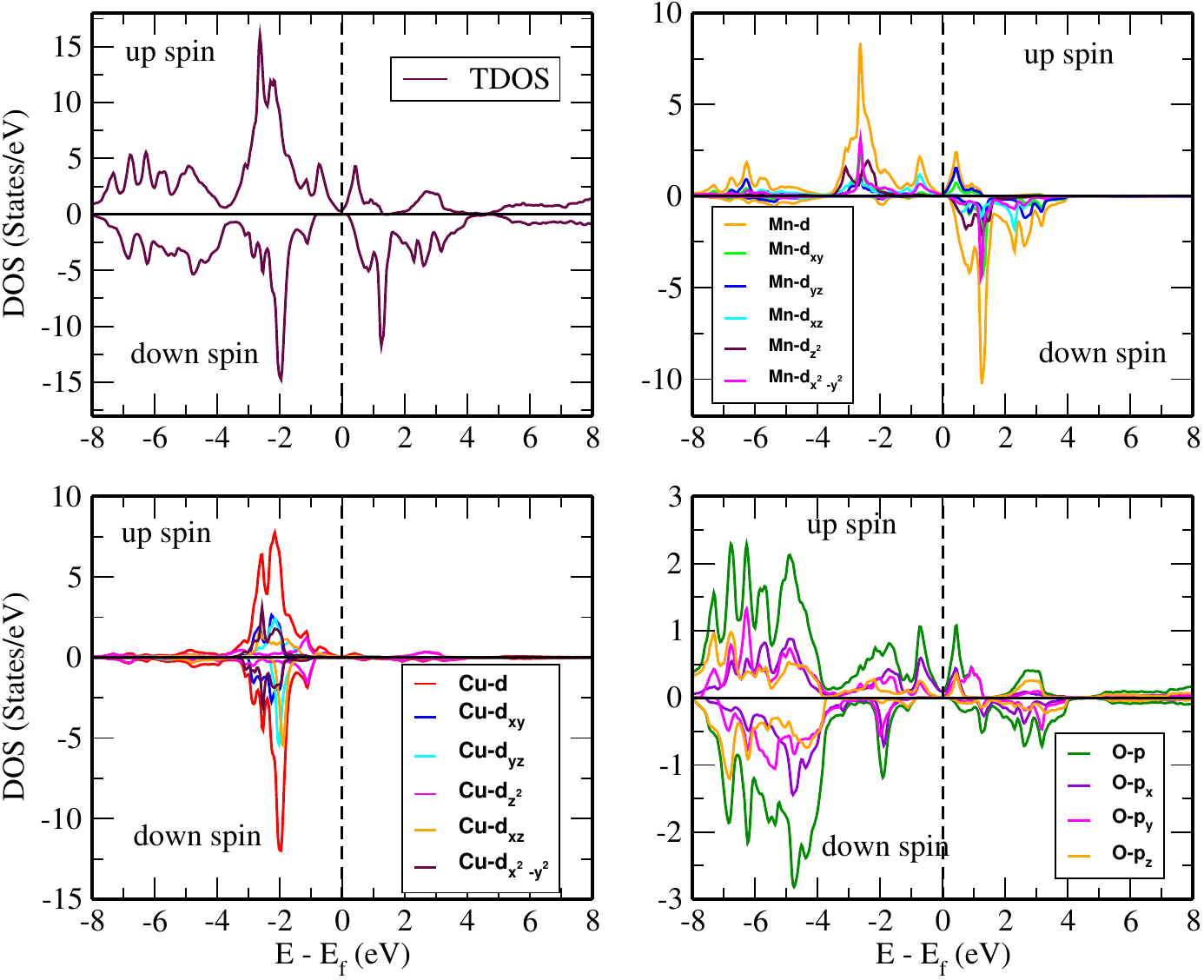}
	\caption{}
	\label{dos1}
\end{subfigure}
\begin{subfigure}[b]{0.5\textwidth}
	\centering
	\includegraphics[width=1.0\textwidth]{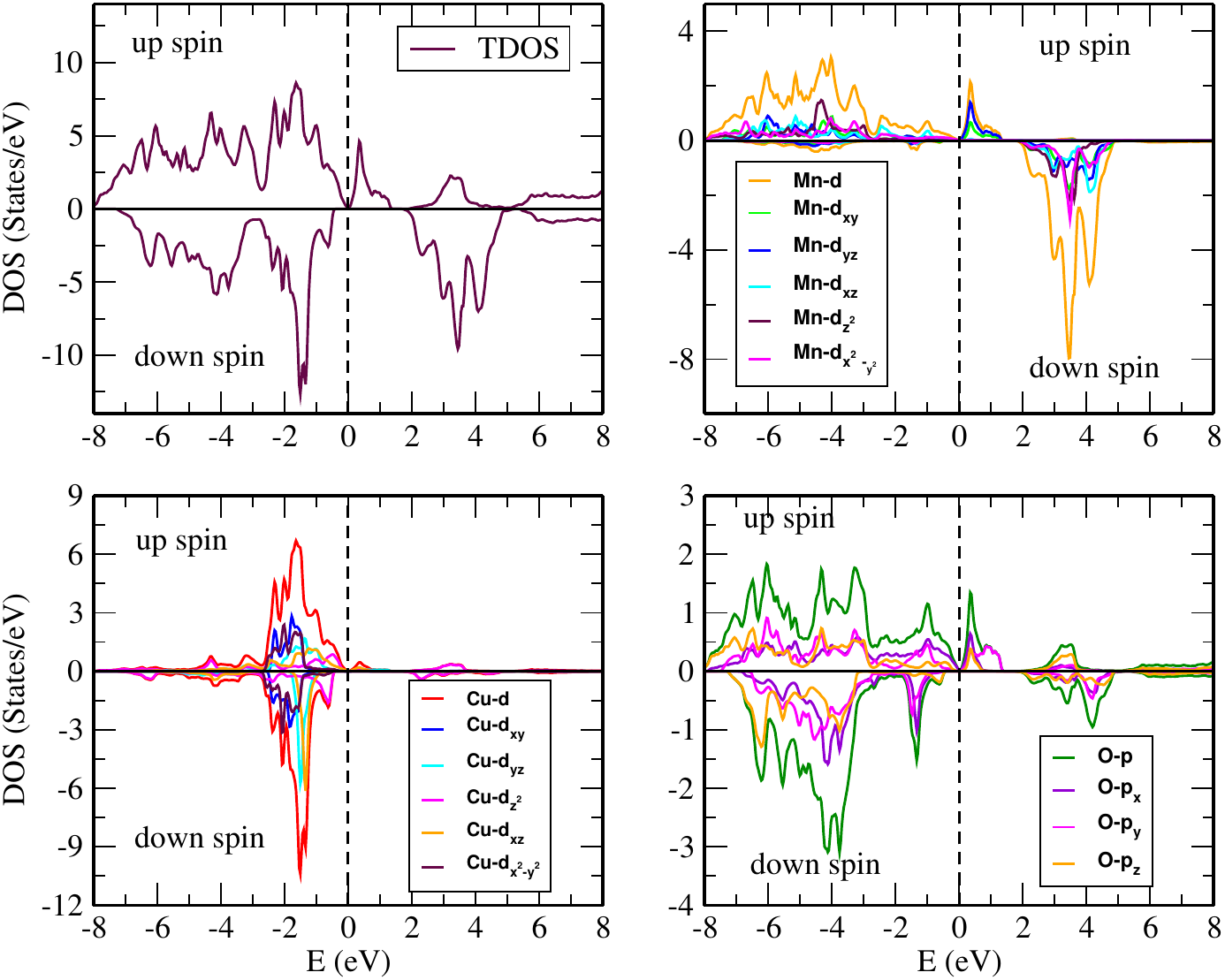}
	\caption{}
	\label{dos2}
\end{subfigure}
	\caption{Total and partial density of states of $\textrm{CuMnO}_{2}$ in the ferromagnetic state, calculated under GGA [panel (a)] and GGA+U [panel (b)] approximations. Detailed orbital contribution 
	of each atom in a given unit cell are also shown.}
        \label{Fig:dosFM}
\end{figure}
\begin{figure}[htbp]
	\centering
	\begin{subfigure}[b]{0.5\textwidth}
	\includegraphics[width=1.0\textwidth]{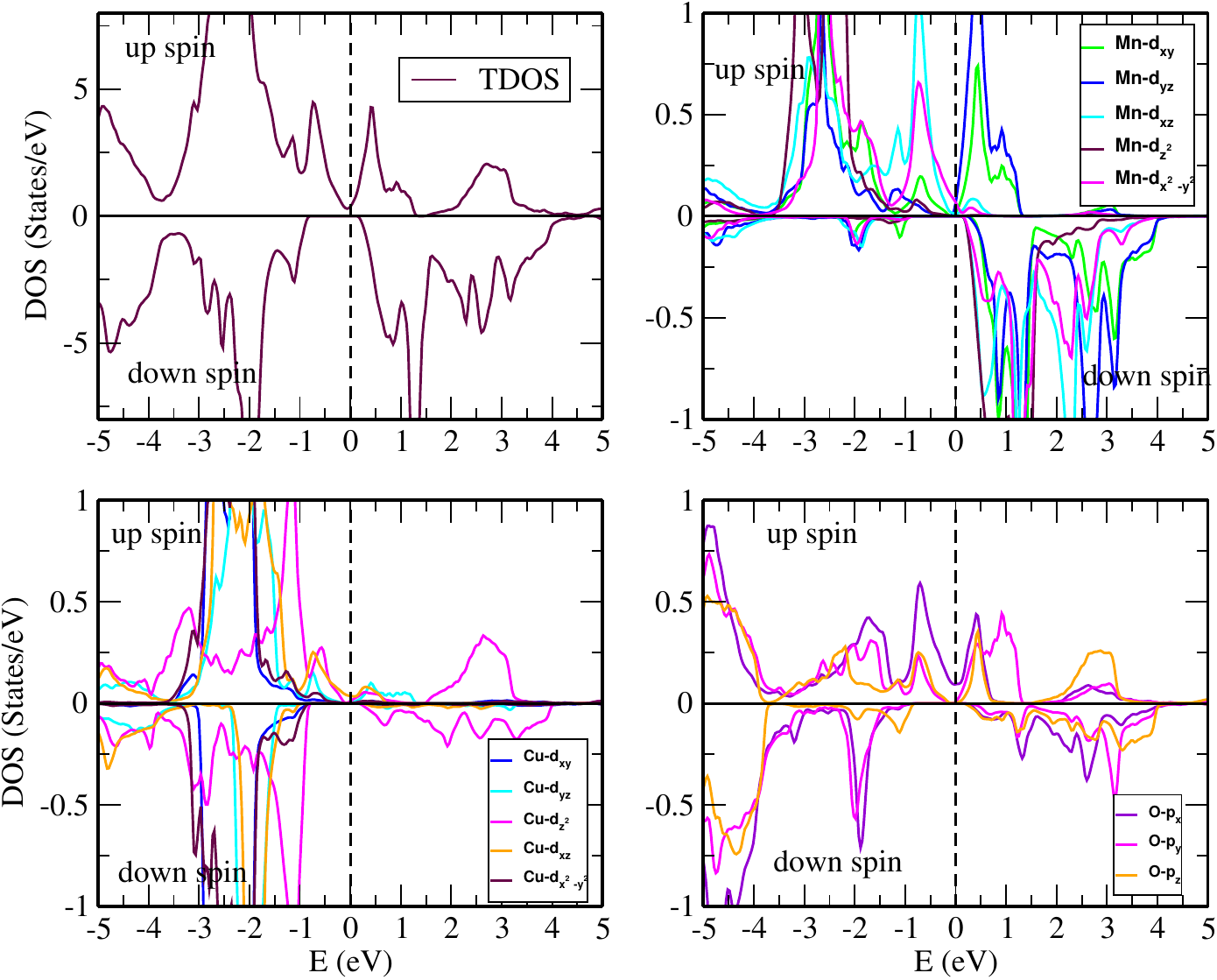}
	\caption{}
	\label{dos3}
\end{subfigure}
\begin{subfigure}[b]{0.5\textwidth}
	\includegraphics[width=1.0\textwidth]{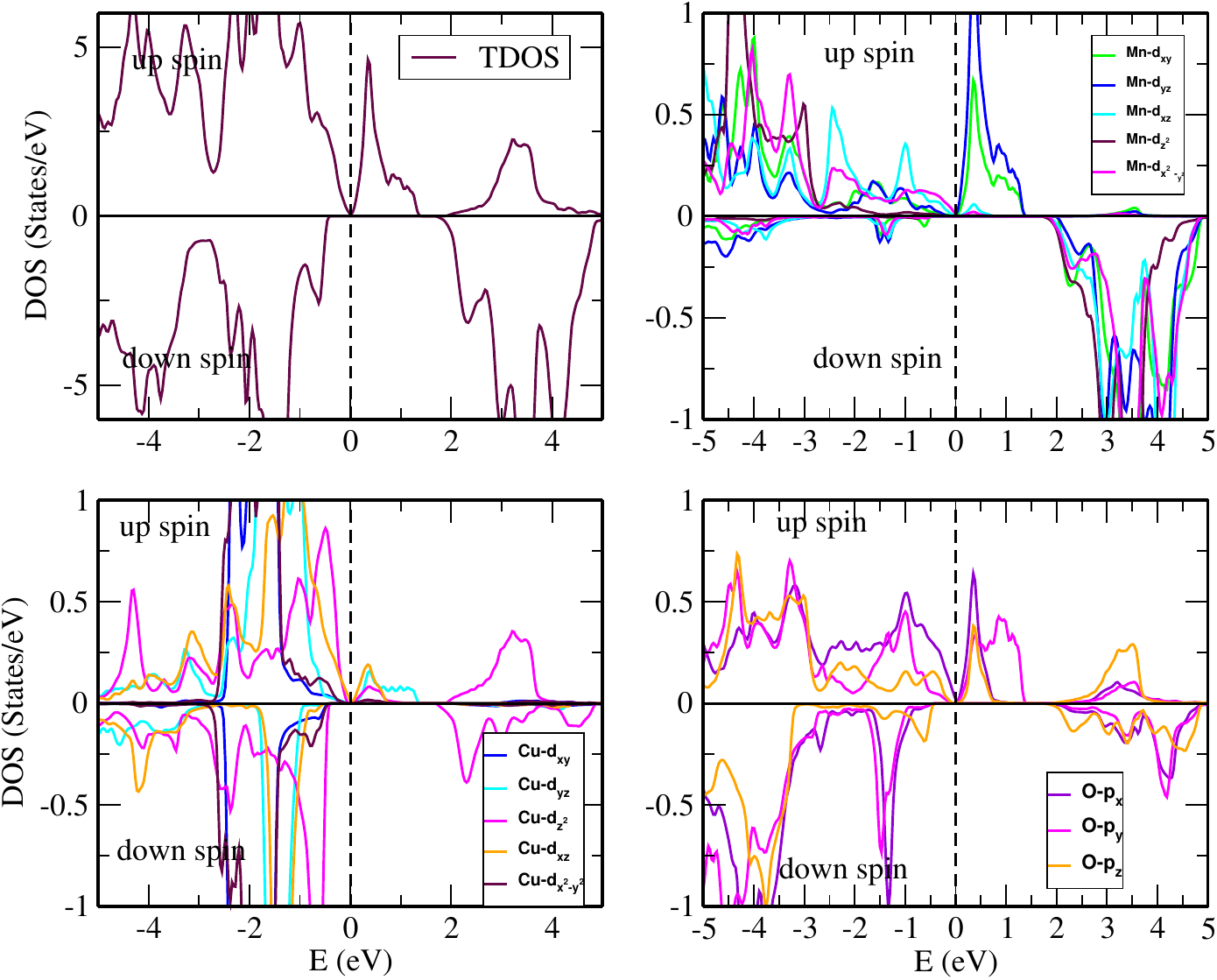}
	\caption{}
	\label{dos4}
\end{subfigure}
	\caption{Same plot as in Fig~\ref{Fig:dosFM} but in a reduced energy window. Reduced energy window has been chosen to enhance the clarity of orbital resolved partial DOS near the Fermi level.}
        \label{Fig:dosFMreduce}
\end{figure}

In Fig.~\ref{Fig:dosFM} we have plotted total as well as partial density of states (pDOS) for various atoms under GGA [panel (a)] and GGA+U [panel (b)] approximations. We also have shown orbital resolved pDOS for 
each atom. A detailed plot in a reduced energy window in Fig.~\ref{Fig:dosFMreduce} shows that the density of states for the up spin channel under GGA approximation [panel (a)] is small but finite while the same under GGA+U 
approximation [panel (b)] vanishes at the Fermi level. The density of states in both the cases shows asymmetric power law behaviour at the Fermi level which is a characteristics of the half-metallic state. The density of 
states for the down spin channel shows a clear gap at the Fermi level under both GGA and GGA+U approximation. The power law behaviour of up spin density. The up spin density of states arises due hybridization between Mn-$3d$ 
and O-$2p$ orbitals. In particular Mn-$d_{x^2-y^2}$ and O-$2p_{x}$ hybridized orbital makes dominant contribution to density of states at the Fermi level. This p-d hybridization plays vital role to stabilize the half-metallic
ferromagnetism in $\textrm{CuMnO}_{2}$. In addition to the above, one can see that Cu-$3d$ states have negligible contribution in conduction as well as valence band to the total DOS for both the spin channels near Fermi level. This p-d hybridization plays vital role to stabilize the half-metallic ferromagnetism in $\textrm{CuMnO}_{2}$.
 
 \section{Magnetic Properties}
 We also have calculated magnetic moment per unit cell as well as magnetic moment of individual atoms in both the phases. In Tab.~\ref{Tab:MagMom} we have shown the total magnetic moment per unit cell ($\textrm{M}_{tot}$) 
 and the magnetic moments contributed by the individual atom for both the ferromagnetic and antiferromagnetic state of $\textrm{CuMnO}_{2}$. As can be clearly seen the main contribution of total magnetic 
 moment is due to the Mn atoms whereas the induced magnetic moments on Cu and O atoms are negligibly small. As already mentioned due to crystal field effects each $Mn^{+3}$ ion is in $t_{2g}^{3}e_{g}^{1}$ configuration. 
 The electrons in 3 fold degenerate $t_{2g}$ orbital forms a core spin with $S=3/2$ and magnetic moment $3\mu_{B}$. The electron in $e_{g}$ orbital hybridizes with neighbouring $O$ atoms. Also two layers of $MnO_{6}$ 
 octahedra are connected via O-Cu-O bond along the $c$-axis. Interestingly, in the ferromagnetic state total induced moment on 6 oxygen atom and a Cu atom is exactly equal to the missing 0.301$\mu_{B}$ moment for each $e_{g}$  electrons. The calculated magnetic moments are very close to the reported values~\cite{}. 
\begin{table}[htbp]
	\centering
	\caption{Calculated magnetic moment of individual atom and total magnetic moment of  $\textrm{CuMnO}_{2}$.}\label{Tab:MagMom}
	\begin{tabular}{c|c|c|c|c}
	\hline
	Systems & Cu ($\mu_{B}$) & Mn ($\mu_{B}$) & O ($\mu_{B}$) & $\textrm{M}_{tot}$ ($\mu_{B}$) \\
	\hline
	$\textrm{CuMnO}_{2}$ (FM) & 0.019 & 3.699 & 0.047 & 7.624 \\
	\hline
	$\textrm{CuMnO}_{2}$ (AFM) & 0.016 & 3.574 & 0.024 & 0.0 \\
	\hline
	\end{tabular}
	
\end{table}
\section{Conclusions}
To summarize, we have calculated the electronic band structures and density of states of $\textrm{CuMnO}_{2}$ in both the ferromagnetic and antiferromanetic states under GGA and GGA+U approximation. The lattice parameters 
obtained from relaxed structures show good agreement with other theoretical and experimental results. The electronic band structure in the antiferromagnetic state shows semiconducting behaviour with indirect band gap - 
consistent with experimental observations. However the calculated band gap ($\sim 0.5$ eV) overestimates the experimentally observed value. This is due to the inadequacy of the exchange-correlation function used under 
GGA approximation. The orbital resolved partial density of states shows that the valence band arises predominantly due to hybridization between Mn-$3d_{x^2-y^2}$ and O-$2p_{x}$ orbitals, whereas the conduction bands arises 
due to hybridized Mn-$3d$ and O-$2p$ states. The broad peaks in the conduction band arise due to transition between crystal field split Mn-$3d$ orbitals. Cu-$3d$ orbitals have minor contribution to valance bands and 
practically no contribution to conduction bands.

  The electronic band structure in the ferromagnetic state shows band gap in the down spin channel whereas the bands in the up spin channel touches the Fermi level. This behavior is consistent with half-metallic state with 
100\% spin polarization in the ferromagnetic phase. The half metallicity arises due to \textit{double exchange} mechanism in which the $e_{g}$ electron spin is strongly coupled to the $t_{2g}$ core spin through Hund's 
coupling, $J_{H}$. This facilitates conduction of up spin electrons while the conduction down spin electrons costs an energy $J_{H}$. As a result the energy bands for the up spin channel is gap less while that for the down 
spin bands are gapped. Due to this unique behavior, $\textrm{CuMnO}_{2}$ in the ferromagnetic state has a potential application as a spin injector for spin valve in spintronic devices. The density of states near the 
Fermi level shows asymmetric power law behaviour. From the orbital resolved partial density of states we infer that hybridized Mn-$d_{x^2-y^2}$ and O-$2p_{x}$ orbital gives rise to bands near the Fermi level.

We also have calculated the total magnetic moment for a given unit cell as well as magnetic moment of individual atoms in the both the antiferromagnetic and ferromagnetic state. In the ferromagnetic state the magnetic moment 
of Mn atoms $3.699\mu_{B}$ arises due to localised $t_{2g}$ electrons (3$\mu_{B}$) and itinerant $e_{g}$ electrons while the induced magnetic moment on O and Cu atoms arises due to hybridization on $e_{g}$ electrons with 
O-$2p$ orbitals along the Mn-O-Mn bond and hybridization between O-$2p$ and Cu-$3d$ orbitals along the O-Cu-O bonds connecting two layers of $MnO_{6}$ octahedra.

\section{Acknowledgements}
We would like to thank Kazi Nazrul University for partial research support. One of us (A.S) would like to thank UGC, Government of India for providing financial support. One of us (N.P) would like to thank IIT, Kharagpur for providing hospitality where part of the manuscript was written.
\bibliographystyle{unsrt}
\bibliography{reference}
\end{document}